\documentclass[a4paper,11pt]{article}
\usepackage{pos}

\title{Radiation pressure instability: from heart-beat states in black hole binary systems
to Quasars and Changing-Look AGN}
\ShortTitle{Radiation pressure instability}

\author*[a]{Agnieszka Janiuk}
\author[a]{Bozena Czerny}
\author[a]{Pulkit Ojha}
\author[b]{Yuri Cavecchi}
\author[c,d]{Federico Vincentelli}

\affiliation[a]{Center for Theoretical Physics,
  Al. Lotnikow 32/46, 02-668, Warsaw, Poland}
\affiliation[b]{Departament de Fis´ıca, EEBE, Universitat Polit`ecnica de Catalunya, Av. Eduard Maristany 16, 08019 Barcelona, Spain}
\affiliation[c]{Fluid and Complex Systems Centre, Coventry University, CV1 5FB, UK}
\affiliation[d]{INAF—Istituto di Astrofisica e Planetologia Spaziali, Via del Fosso del Cavaliere 100, I-00133 Roma, Italy}

\emailAdd{agnes@cft.edu.pl}
\abstract{Radiation-pressure instability was identified soon after the seminal classical accretion disk models of Shakura \& Sunyaev and Novikov \& Thorne, yet its full implications remain an active area of investigation. These models form the backbone of our understanding of accretion onto compact objects and successfully describe the phenomenology of black hole and neutron star X-ray binaries, as well as luminous active galactic nuclei (AGN), in the regime of high mass accretion rates. At luminosities approaching a significant fraction of the Eddington limit ($L/L_{\rm Edd} \ge 0.1$), standard thin disks are predicted to become thermally unstable due to the dominance of radiation pressure. This prediction has found empirical support in several Galactic stellar-mass black hole systems, where the instability manifests as quasi-periodic, large-amplitude luminosity oscillations, so-called “heartbeat states”, and has been proposed as a driver of observed signatures of deterministic chaos in accretion-driven light curves.

The scope of radiation-pressure-induced variability extends beyond stellar-mass black holes: both black holes across mass scales and accreting neutron stars can exhibit related behavior, though the presence of a boundary layer in neutron stars adds complexity and offers a unique laboratory for testing the interplay between accretion dynamics and the central object. On extragalactic scales, the instability has been invoked to explain the duty cycles and apparent short lifetimes of radio-loud AGN, as well as the dramatic spectral-state transitions seen in Changing-Look AGN.

Despite these advances, the theoretical picture remains unclear. State-of-the-art magnetohydrodynamic (MHD) simulations have not yet reached consensus on the conditions under which the radiation-pressure instability fully develops or is suppressed. Magnetic fields, which mediate angular momentum transport and couple to both gas and radiation, emerge as a potentially critical factor in modulating the instability’s onset and nonlinear evolution. This lecture will revisit the historical development of the instability concept, survey the current observational and theoretical status across different classes of accreting compact objects, and highlight the key open questions
in the theory
and the connection to observed variability phenomenology, that will shape the next fifty years of disk instability research.
}

\FullConference{
}

\begin{document}
\maketitle

\section{Introduction}

Accretion disks are known to power the X-ray binaries with compact objects and AGN.
The variability of observed emission  encodes
physical processes underlying the heating and cooling of the plasma and the transport of angular momentum that enables accretion. The stochastic type of variability is found in most states of various objects, across their mass scale. However, a distinct type of periodic, or quasi-periodic flares, is occasionally observed, pointing towards a deterministic nature of the process.

The physical mechanism behind such cyclic variations is the Radiation Pressure Instability (RPI). This instability produces regular outbursts, and flare-like events that represent low-dimensional chaos. It may also be responsible for the duty cycles in some objects and lead to an on-off behavior or intermittent activity. 
The physical nature of the RPI driven variability is connecting the micro- and macro-scales, because it is related to the stress tensor scaling with pressure in the plasma, and also to directly probe the viscosity magnitude and timescales of secular changes. 
Observational signatures of the RPI span from the heartbeat states observed in microquasars (tens-hundreds of seconds) to Quasars and Changing-Look AGN (tens -thousands of years).

\section{Brief history of accretion disks}

  \begin{figure}
  \centering
    \includegraphics[width=\textwidth]{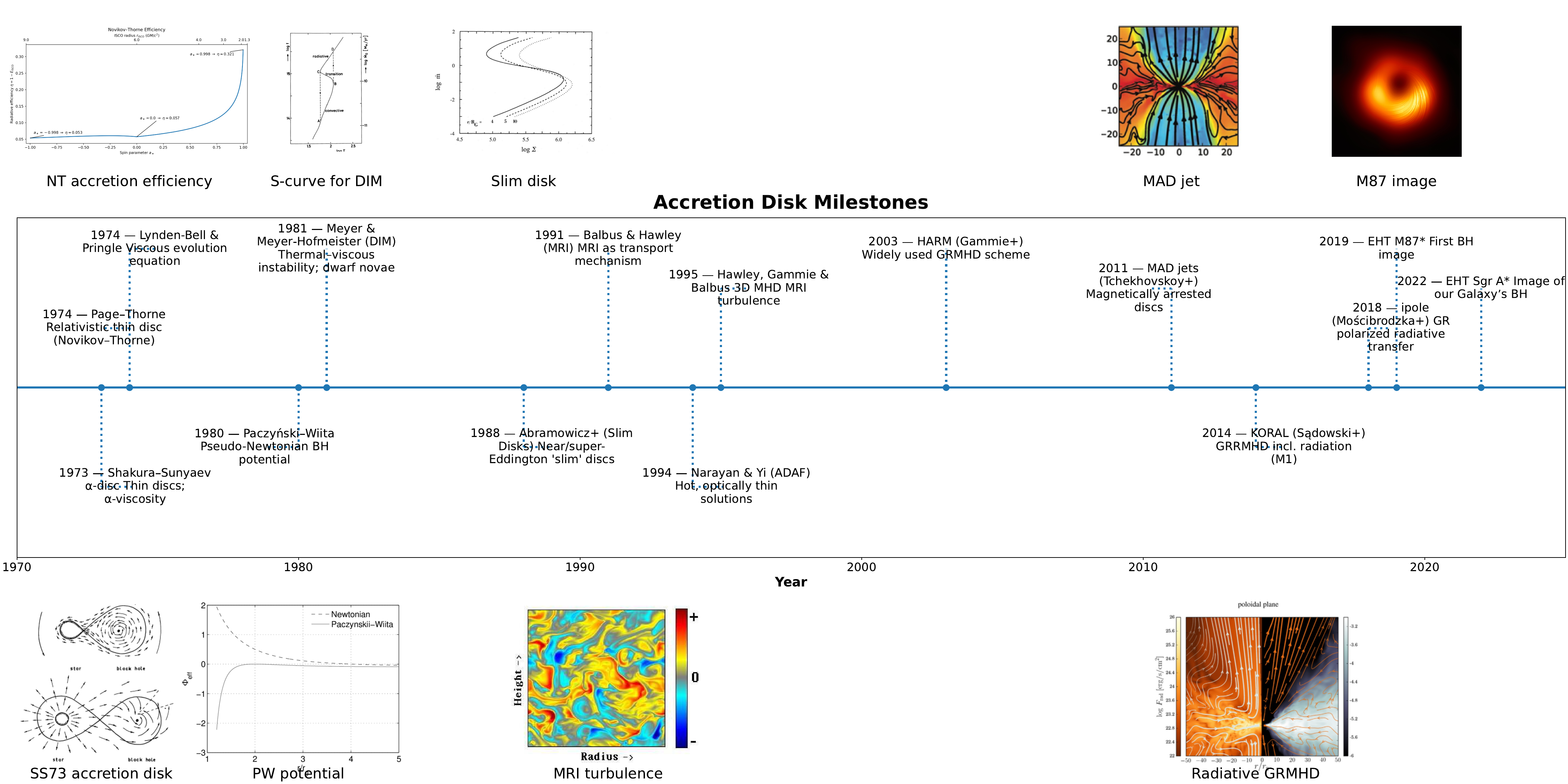}
    \caption{Schematic timeline showing milestones in the accretion theory}
    \label{fig:fig1}
  \end{figure}

The theory of accretion disks was proposed to explain the appearance of high energy emitting sources related to white dwarfs, black holes or neutron stars residing in binary systems located in our Galaxy. These compact stars gain material from their companion stars in either high mass or low mass regime, where the companion star may fill its Roche lobe and donate material through the Lagrangian $L_{1}$ point, or via the focused stellar winds. N. Shakura and R. Sunyaev proposed in  \cite{ss73} that the material in both cases forms a geometrically thin, optically thick disk in the orbital plane of the binary system. In order to loose the angular momentum and proceed to lower orbits around the accretor, the gas layers shear against each other and the angular momentum is transported outwards by viscosity. The turbulent nature of the process implies an effective scaling of viscous turbulent velocity and length-scale, with the speed of sound and disk vertical scale-height.
A concurrent model proposed by Novikov \& Thorne \citep{NT1973} proved the efficiency of the process leading to viscous energy dissipation being very large with respect to the rest-mass energy of accreting particles, and growing enormously with the black holes spin. 
Hence, accreting Kerr black holes are the most powerful sources of energy expected so far in the Universe.

In Figure \ref{fig:fig1}, we present a timeline visualising the milestones achieved in the accretion disk theory. For completeness, the graph ends with some current state-of-the art numerical models \footnote{We mention Narayan-Yi ADAF model, the GR MHD numerical schemes, MAD disks and relativistic jets, and polarized radiation transport simulations as the milestones}, which are not the scope of this Proceeding, but will be commented briefly in the last Section.
In 1970's along with Shakura-Sunyaev model, Lynden-Bell \& Pringle \cite{LBP1973}
showed that a thin Keplerian disc with an effective kinematic viscosity evolves diffusively. The viscosity causes angular momentum to flow outward and mass to flow inward. They derived the surface density diffusion equation with self-similar solutions, and for a viscosity that scales as a power of radius, their explicit similarity solutions describe how an initially ring-like mass distribution spreads over time. 
This work provides a general framework for the time-dependent evolution of viscous accretion discs.

The viscous and thermal instability mechanism that can potentially lead to the disk disruption, was identified by Lightman \& Eardley \cite{lightman74} who showed that the standard Shakura–Sunyaev $\alpha$-disk with stress proportional to the total pressure is unstable in the radiation-pressure-dominated inner region. 
The equilibrium branch of this solution exhibits (i) thermal instability, because the heating grows faster than cooling for a temperature perturbation, and (ii) viscous (secular) instability, because the effective diffusion coefficient becomes negative, since accretion rate decreases with increasing surface density.
The thermal instability itself was then discussed in the context of two-temperature optically thin disks by %Shapiro, Lightman \& Eardley 
\cite{SLE76}.
These SLE disks are unstable because of thermal runaway: their heating rate grows faster than cooling rate, at fixed density (however, they are viscously stable).

A next milestone that boosted the first numerical works on accretion disks evolution, was the Paczynski-Wiita potential, introduced in \cite{PW1980}. It exactly reproduces the locations of the innermost stable circular orbit (ISCO) and the marginally bound orbit, and gives the correct functional form of the specific angular momentum distribution. Despite the mathematical simplicity, it captures key strong-gravity features of a Schwarzschild black hole. Therefore, while still working in a Newtonian hydrodynamics framework, it is widely used in analytic models and simulations of black-hole accretion disks when full GR is unnecessary or too expensive.

The disk instability (DIM) model proposed for outbursts driven by a thermal–viscous instability, is not a purely thermal one. The mechanism was discussed in \cite{meyers} and \cite{Smak84}.
They showed that locally, the instability is triggered when the disc temperature enters the hydrogen partial-ionization regime, because then the opacity and pressure respond very steeply to temperature, so a small heating makes heating rate rise faster than cooling. 
In an $\alpha$-disc, the viscosity also depends strongly on temperature. Along the middle branch of the S-curve 
(visualised in the insert of Fig. \ref{fig:fig1})
the solution is both thermally and viscously unstable, 
so the same physical mechanism gives thermal and secular (viscous) instability. This is producing the limit cycle between the cold and hot branches, commonly used to explain the outbursts of Dwarf Novae.
\footnote{Note that some reviews (e.g. Lasota 2001, Hameury 2009, 2019) explicitly trace the thermal–viscous DIM back to Hoshi (1979) for identification of the instability.
 The work of Osaki (1974) is also often cited as the first explicit accretion disk outburst model for dwarf novae (the seed of the DIM), while Hoshi (1979) provided the detailed thermal instability and bimodal-state disc structure (the S-curve). }

The RPI instability is also of thermal-viscous type, as argued above. In order to produce an outburst cycle, some stabilizing mechanism must be found. It was possible only thanks to the slim disk model  \cite{abramowicz1988} who presented a new branch of stationary accretion-disk equilibria at $\dot M \sim \dot M_{\rm Edd}$ around black holes.  This branch closes the S-curve of RPI instability (cf. Fig. \ref{fig:fig1}). It exists once the radial advection of heat in an optically thick, nearly Keplerian disk is properly included.
These “slim disks” are geometrically thicker than standard thin disks but they are not as torus-like as the other famous solutions - Polish “doughnuts”, e.g \cite{PW1980}. A substantial fraction of the dissipated energy is advected inward rather than radiated locally, so their luminosity saturates and can only modestly exceed the Eddington value. The slim disks are quasi Keplerian, and keep the $H/R$ ratio well below 1.

\section{Physical origin of Radiation Pressure instability}

  \begin{figure}
  \centering
  \includegraphics[width=\textwidth]{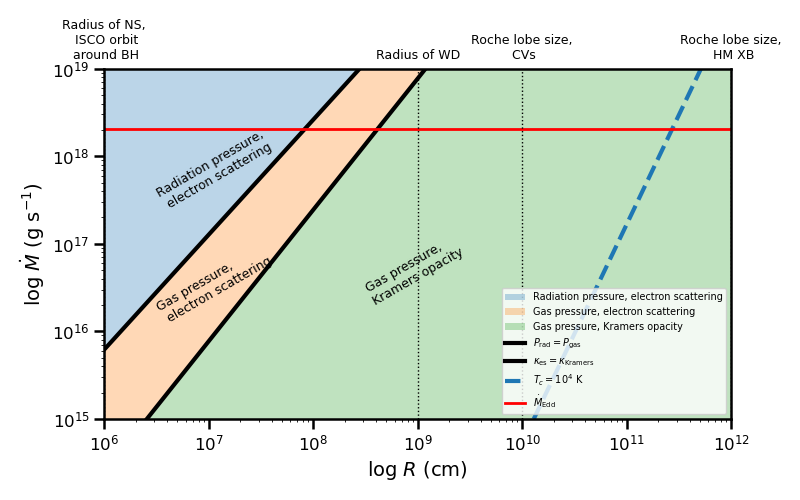}
    \caption{Regions of accretion disks dominated by different physical processes affecting thermal balance: pressure components, opacity sources. The disk outer radius must fit into the Roche lobe size, depending on the typical mass ratio of the binary. The inner radius can extend only to the ISCO orbit for BH, but cannot be smaller than WD radius for Cataclysmic Variables. The red solid line marks Eddington accretion rate limit. The blue dashed line marks region with temperature $10^{4} K$ of Hydrogen recombination, where the other opacity sources are important. }
    \label{fig:zones}
  \end{figure}

In Fig. \ref{fig:zones} we show the physical regions in the Shakura-Sunyaev $\alpha$ disks where either radiation pressure and electron scattering opacity dominates (blue region), or the gas pressure with electron scattering 
is dominant (orange region) or the gas pressure with Kramer's opacity dominates (green region). 
The location of the zones depends on the accretion rate. The Eddington accretion rate for adopted compact object with mass $1.4 M_{\odot}$ (black hole or neutron star) and accretion efficiency of $\eta=0.1$ is marked with red horizontal line. 
The location of the radiation pressure dominated zone is plotted for assumed viscosity parameter $\alpha=0.1$.
In addition, the blue line marks the region where the disk midplane %(central) 
temperature in the disk drops below $10^{4}$~K, so that other opacity sources than Kramers' become important, due to hydrogen recombination. That region is unstable, according to the DIM model, until temperature drops below $\sim 3\times 10^{3}$~K, when the hydrogen is fully neutral. 

We notice that for WD, BH and NS, both regions of instabilities due to $P_{\rm rad}$ and Hydrogen partial-ionisation can exist in the same disk, if only the accretion rate is high and the disk size, limited by Roche lobe size, is also large enough.
For WD, the radiation pressure zone is not extending beyond the WD radius, so in practice we will not observe this kind of instability.

\begin{figure}
  \centering
  \includegraphics[width=\textwidth]{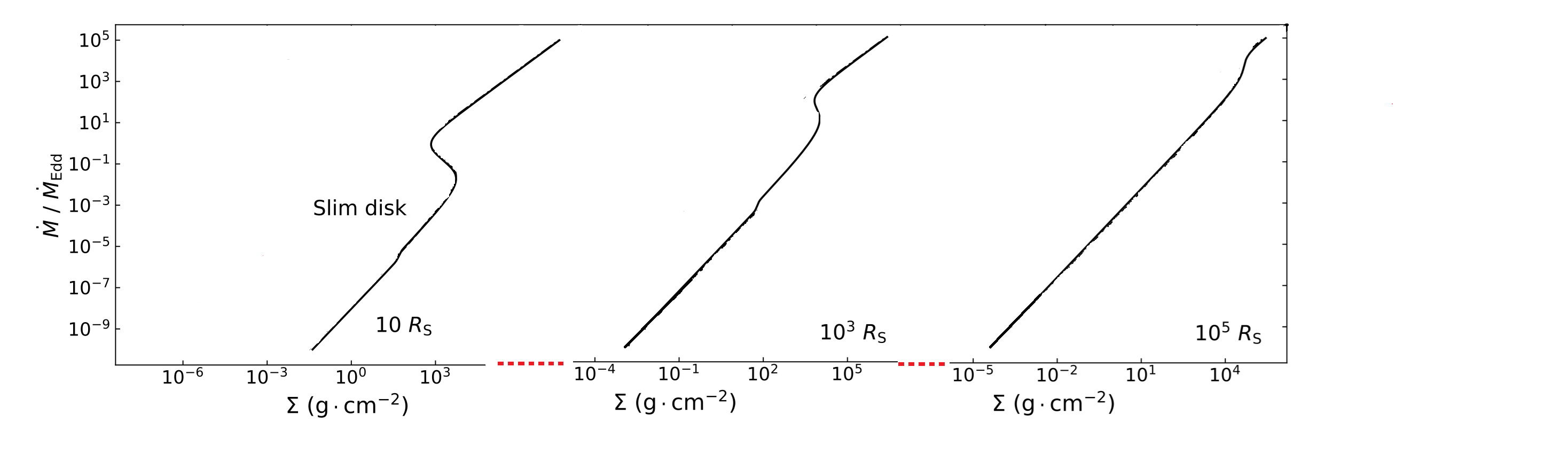}
  \caption{Symbolic representation of the  thermal equilibria (S-curves) of the accretion disk.
   Since the curve apparently shifts up with
  radius, at high $\dot M$ the instability might be driven by more distant, rather than inner, parts. However, the shape of the curve also varies with the radius, and at the largest radius showed above ($10^{5} R_{S}$) the instability disappears, so the outermost part  of the flow is always stable against RPI instability. Presented plots are based on the recent computations by \cite{liu}.}
    \label{fig:s_curve}
  \end{figure}

Conceptually, in the plot of local thermal equilibria 
(see Fig. \ref{fig:s_curve}; 
at fixed radii), the inclusion of advection gives an extended S-curve. A lower Shakura–Sunyaev thin-disk branch, an intermediate (unstable) radiation-pressure region, and an upper, advective, slim-disk branch that is thermally and viscously stable. 
This is closely analogous in mathematical structure to the DIM S-curve for dwarf novae, even though the physical origin of the extra branches (radiation pressure + advection rather than H-ionization) is different.

The local equilibrium solutions are found by solving the heating and cooling balance. For slim disks, the relevant terms, describing the (vertically averaged) viscous heating and radiative plus advective cooling, are as follows:
    \begin{equation}
Q_{+} = \frac{3}{2} \alpha P H \Omega_{\rm K}; ~~~~
Q_{-} = \frac{4 \sigma_{\rm B} T^{4}}{3 \kappa \Sigma}+Q_{\rm adv}; ~~~~
Q_{\rm adv} = - \frac{\dot M}{2 \pi R^{2}}{P\over \rho} 
\nonumber
\end{equation}
Here, $P$ denotes the total (gas plus radiation) pressure, $\Omega_{\rm K}$ is the Keplerian angular velocity at radius $R$, $\Sigma$ is the disk surface density at this radius, $H$ is the disk thickness, and $T$ is its mid-plane temperature. The opacity is dominated by electron scattering, $\kappa=\kappa_{\rm es}$.
The local solutions form the S-curve \cite{abramowicz1988}, which for super-Eddington rate actually have a straight shape,
as shown in Figure \ref{fig:s_curve}.

To analyze the stability of thermal equilibria, one may 
study small perturbations of density and temperature over time.
In the equilibrium, the following balance is satisfied:

    \begin{equation}
      \frac{\partial T}{\partial t} = Q^{+}-Q^{-} = h(T,\Sigma)
      \nonumber
    \end{equation}
    together with the diffusion equation for surface density:
    
    \begin{equation}
      \frac{\partial \Sigma}{\partial t} = (t_{\rm th}/t_{\rm visc}) k(T,\Sigma)
      \nonumber
    \end{equation}
 where the timescales are defined as 
 $t_{\rm visc}=R^{2}/\nu$; ~~ $t_{\rm th}=({v_{\phi}\over c_{\rm s}})^{-2} t_{\rm visc}$, with $\nu$ being a kinematic viscosity, and $c_{\rm s}$ the speed of sound.

\begin{figure}
  \centering
  \includegraphics[width=0.45\textwidth]{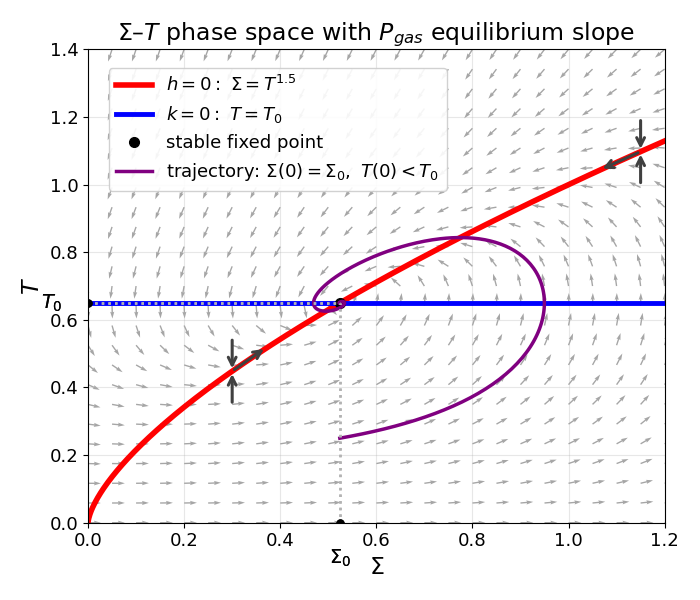}
    \includegraphics[width=0.45\textwidth]{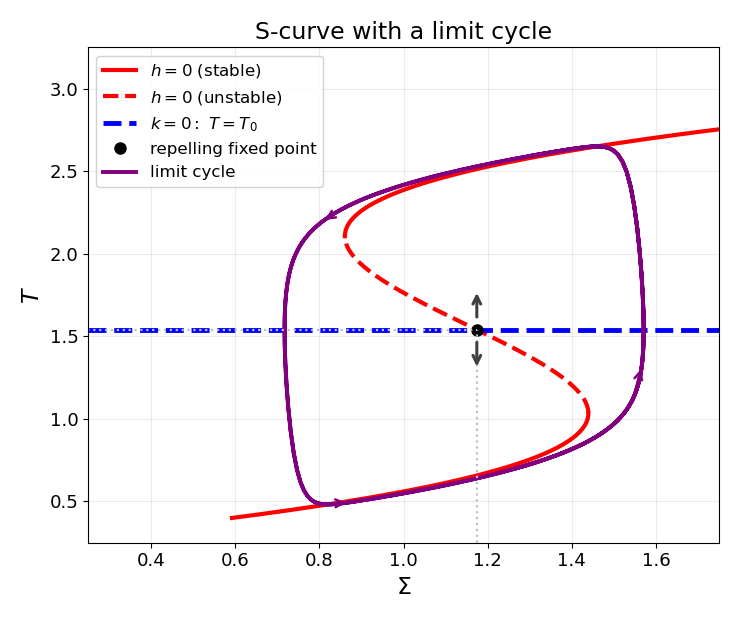}
  \caption{Attractor point on the stable branch of disk solutions (left) and limit cycle around repelling fixed point in the unstable branch of disk solutions (right).
  }
  \label{fig:attractor}
  \end{figure}

 Nature of the stable and unstable fixed points (attractor and repeller) is visualized in Fig. \ref{fig:attractor}.
The solution is a fixed point that lies on the intersection
of $h$ and $k$ curves, i.e. $Q^{+}=Q^{-}$ and $T=T_{0}$ are satisfied simultaneously. On the stable branch of an S-curve, if $\partial \Sigma /\partial T> 0.$, the fixed point is attracting, i.e. any small perturbation of temperature is damped and the solution comes back to this fixed point. If the temperature rises slightly, the system moves along the $h=0$ line, on the thermal timescale, before it finds the new value of $\Sigma$ (that evolves on viscous timescale), searching for the new $k=0$ intersection. But then the accretion rate will increase from local one $\dot M_{0}$ corresponding to the temperature $T=T_{0}$, so the local density will immediately drop. The temperature drops then accordingly, and the perturbation is damped. 

A reverse situation is found at the unstable branch, with $\partial \Sigma / \partial T <0$, where the fixed point is repelling. Here, any small perturbation of temperature leads to a runaway heating, with surface density decreasing on viscous timescale and accretion rate increasing above $\dot M_{0}$. This thermal-viscous instability may be eventually controlled, if only another stable branch of equilibria is found. This stabilizing branch is the upper part of the S-curve. There again, a small perturbation in temperature, ie. its decrease, will lead to runaway cooling, as the local accretion rate will drop below $\dot M_{0}$. 
As a result, for the accretion rates in the range corresponding to the negative slope of the S-curve, the disk annulus would oscillate between the lower and upper stable branches, and a type of limit-cycle oscillation is produced. 
    
\section{Numerical solutions}

The local stationary solutions were discussed above, as resulting from the thermal heating and cooling balance.
The closing equation which enable to find global structure of an accretion disk for a given accretion rate, and parameterized with viscosity $\alpha$ and black hole mass $M$, is:
\begin{equation}
F_{\rm tot} = {3 G M \dot M \over 8 \pi R^{3}}
\nonumber
\end{equation}
which gives the rate of locally dissipated heat due to gravitational potential energy of the black hole, and must be equal to the $\alpha$ disk viscous heating term. 
The model must also be supplemented with hydrostatic balance, of the form $c_{s}^{2}= P_{\rm tot}/\rho = H^{2} \Omega^{2}$
that relates speed of sound with rotation.
Solving the set $Q_{+}=Q_{-}=F_{\rm tot}$ for temperature and density, at every radius, enables to find a global structure of the disk.

\subsection{Global time dependent solutions}

    The 1D time dependent model proposed by \cite{janiuk2002} is solving the evolution of accretion disk subject to RPI. 
 Governing equations are viscous diffusion and energy transport, and they read:
\begin{equation}
\frac{\partial \Sigma}{\partial t} = \frac{1}{r} \frac{\partial }{\partial r}( 3 r^{1/2} \frac{\partial }{\partial r}( r^{1/2} \nu \Sigma )) \nonumber
\end{equation}
and
\begin{equation}
\frac{\partial \ln T}{\partial t} + v_r \frac{\partial \ln T}{\partial r} %\nonumber
= (\frac{4 - 3 \beta }{12 - 10.5 \beta} )(\frac{\partial \ln \Sigma}{\partial t} - \frac{\partial \ln H}{\partial t} + v_r \frac{\partial \ln \Sigma}{\partial r}  )
+ \frac{Q_{+} - Q_{-}}{(12-10.5 \beta)PH} \nonumber
\end{equation}
 where $\beta = P_{\rm gas}/P_{\rm tot}$ and coefficients in brackets can be derived from the ratio of specific heats, in the advection term see \cite{paczynski1981}, \cite{taam1984}, \cite{lasota1991}.
The changes over the thermal and viscous timescales produce long-term limit cycles oscillations., if the mass supply rate, $\dot M_{0}$, is in the unstable regime. 
The extension of instability zone depends on global parameters, the viscosity $\alpha$,  and on the BH mass, as discussed in \cite{JaniukCzerny2011} - for supermassive BHs the zone is larger.

\section{Observational evidence}

There is a growing observational evidence for the RPI instability of accretion disks being in action in a wide range of astrophysical objects. Below we discuss several key examples, present preliminary results on the sources only recently discovered, and provide prospects for further studies. 

\subsection{Stellar mass black holes}

The heartbeat oscillations were first found in GRS 1915+105 \cite{belloni1999}. The source is presenting in some states large amplitude, quasi-periodic variability matching limit-cycle expectations. Timescales and luminosity swings of these flares are consistent with radiation-pressure-driven cycles, as shown in \cite{janiuk2002}.

The other well established candidate is the microquasar IGR J17091, studied thoroughy in \cite{Altamirano2011}.
Its black hole mass $M=6 M_{\odot}$, spin $a=0$, 
accretion rate $\dot M= 0.86-2.12\times 10^{-8} M_{\odot} $~yr$^{-1}$, were found to be sufficient to explain the RPI-driven disk variability. 
The simulations performed by \cite{janiuketal2015} using our 1-D hydro code GLADIS \footnote{github.org/agnieszkajaniuk/GLADIS} proved that 
limit cycle oscillations due to radiation pressure operate in the source, while the disk  might be partially stabilized by wind outflow, as manifested in the spectral energy distributions with broad absorption features, that were weakened during the 'heartbeat' state intervals. %\cite{janiuketal2015}

Further nonlinear diagnostics was made in \cite{sukova2016} by means of the recurrence plots, and attractor reconstruction. The analysis proved evidence for low-dimensional deterministic structure, not pure noise, in the real data of a handful of X-ray sources, that were compared to the surrogate ones. The interpretation via RPI-driven limit cycles was made on the basis of the Recurrence Matix (RM), defined as:
\begin{equation}
\mathbf{R}_{i,j}(\epsilon) = \Theta (\epsilon - \parallel \vec{x}_i - \vec{x}_j \parallel ), \qquad i,j = 1,...,N, \label{RP_def}
\nonumber
\end{equation}
where $\vec{x}_i = \vec{x}(t_i)$ is ($N$) points in phase space,
reconstructed from time series as:
\begin{equation}
\vec{y}(t) = \{x(t),x(t+\Delta t),x(t+2\Delta t),\dots,x(t+(m-1)\Delta t)\}. \nonumber
\end{equation}

In this analysis, the significance of chaos is quantified by the Renyi entropy difference ($\ln K_{2}$) between the surrogate and real data. 
As a result, apart from the two canonical microquasars, GRS1915+105  and IGR17091, traces of non-lienar dynamics were identified in: GX 339-4, XTE J1550-564 and GRO J1655-40.

\subsection{Neutron stars}
Neutron star SWIFT-J1858.6-0814 also shows limit-cycle oscillations \cite{vincentelli2023}.
This discovery gave the first observational evidence for shared instability between stellar mass black holes and neutron star systems.
The multiwavelength observations hinted also on the effects of obscuration, radiation reprocessing and radio jet emission. Necessary condition to understand the source behavior is therefore its behaviour in other wavelengths.

   \begin{figure}
  \centering
    \includegraphics[width=0.8\textwidth]{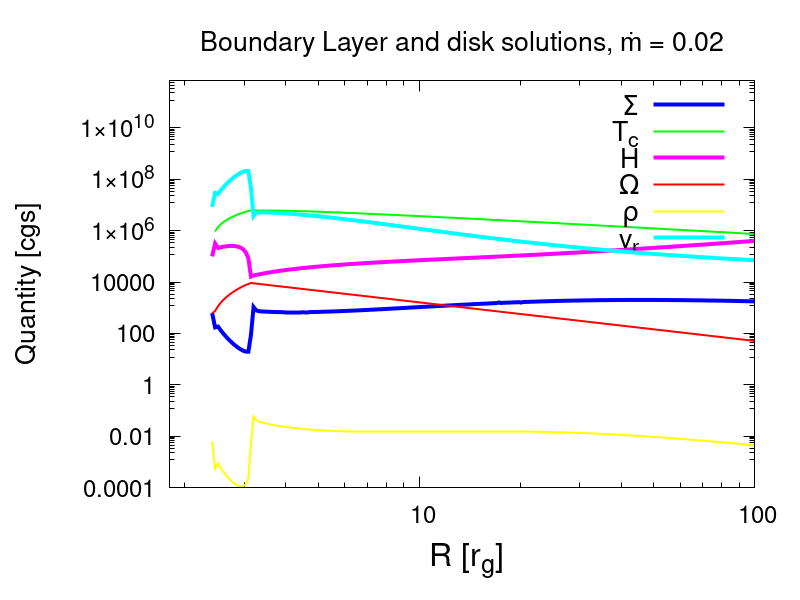}
    \caption{Solutions of the NS boundary layer structure, with a thin Keplerian accretion disk. Quantities plotted are: surface density, central temperature, geometrical thickness, angular velocity, density, and radial velocity. Parameters of the simulation: neutron star mass 
    $M_{\rm NS} = 1.4 M_{\odot}$, neutron star radius $R_{\rm NS}$ = 2.41~$R_{\rm Schw}$, angular momentum coupling parameter $j=1.064$.
    }
    \label{fig:BLsolutions}
    \end{figure}

To model the accretion instability in the neutron star case, we should consider effects of angular momentum transport around the rotating NS, whose hard surface is redistributing the viscously transported energy and decelerates the rotation of the accretion flow. 
As noted by \cite{popham2001}, the radiation flux in the NS boundary layer (BL) is a substantial fraction of the Eddington limiting flux even for luminosities well below ($\sim 0.01$ times) the Eddington luminosity.
The BL layer rotates with sub-Keplerian velocity and is geometrically much thicker than the Keplerian disk surrounding it, hence the irradiation on the disk surface from the BL on the disk may be important, 

To solve for the structure of BL, we follow the approach of \cite{popham2001} and modify the energy balance equations for accretion flow, that now is not assumed Keplerian. We solve for the $\Omega(R)$ profile, along with pressure, temperature and density gradients, with the following set of differential equations.
\begin{equation}
                Q_{+} - Q_{-} = 0 %- Q_{\rm adv} = 0
\end{equation}
where $Q_{-}$
is the vertically averaged radiative cooling, and 
$Q_{+} = Q_{\rm grav,kin} - Q_{\rm trans};$  ~~
with
$Q_{\rm grav,kin} = {{\dot M} \over {4\pi R}} (\Omega_{\rm K}^{2} R + \Omega R^{2} {d\Omega \over dR} + \Omega^{2} R)$ and
$Q_{\rm trans} = - {\dot M \over 4 \pi R} {d \over dR}({\Omega^{2} R^{2} - j \Omega \Omega^{K}_{NS} R_{\rm NS}^{2} })$, standing for heating terms due to gravitational and kinetic energy losses in the BL, and for the divergence of the viscous energy transport rate, as explained in \cite{popham2001}.\\
\noindent
The angular momentum redistribution in the BL reads:
            \begin{equation}
               \frac{d\Omega}{dR} = \frac{v_{\rm r}}{\nu R^2}(\Omega R^2 - j\Omega_{\rm NS}^K R^{2}_{\rm NS}) 
            \end{equation}
and is parameterized by a constant $j$, that may be larger or smaller than 1.0.\\
\noindent
The radial pressure gradient is solved from
\begin{equation}
              v_{\rm r} {d v_{\rm r} \over dR} +  {1 \over \rho} \frac{dP}{dR} = (\Omega^2 - \Omega_{\rm K}^2)R
\end{equation}

The above three equations are solved together, but for simplicity only when $\Omega(R) < \Omega_{K}$. Otherwise, standard Keplerian disk solutions are solved. The resulting profiles of disk structure are shown in Fig. \ref{fig:BLsolutions}.

\subsection{Irradiation of the disk by a central source}

In a realistic situation, the accretion flow in the disk is surrounded by a hot medium, in a form of either a corona, or a spherical source. This medium may interact with the disk and influence the instability pattern in various ways.
Dynamical interaction has been taken into account by \cite{janiuk2005} 
who introduced mass exchange between the disk and the hot corona, under the slab (i.e. 'sandwich-like') geometry. The hot corona above the disk had a virial temperature and varied in time according to the disk instability cycles. Observable consequences of this effect were manifested in time delays between hard and soft X-ray emission, and matched well with observations of GRS 1915+105.

We can also consider the effects of irradiation of the accretion disk from the central source with hard photons, under an
arbitrary geometry. 
The formalism and equations were first introduced by \cite{Dubus1999} who modeled the CV type of instabilities and studied irradiated disks. 

\begin{figure}
  \centering
    \includegraphics[width=0.7\textwidth]{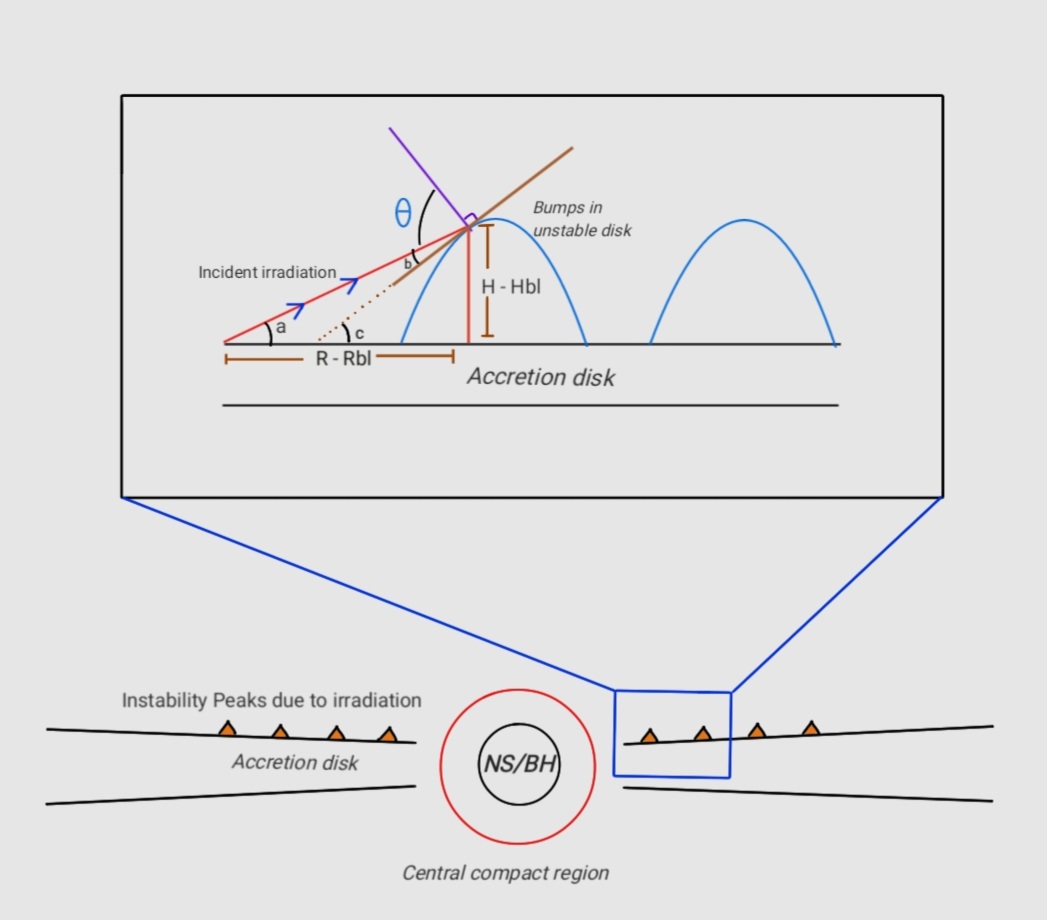}
    \caption{Artistic impression of a disk undergoing radiation pressure instability.
  In the top panel we show how the effect of irradiation from a central
illuminating source (a compact corona or a boundary layer, at position
$R_{\rm BL}$, $H_{\rm BL}$) will affect only the inner part of the instability waves.
The effect of irradiation is proportional to the cosine of the angle
$\theta$ between the incident ray and the normal to the surface of the disc.}
    \label{fig:geom_irr}
  \end{figure}

\begin{figure}
  \centering
    \includegraphics[width=0.8\textwidth]{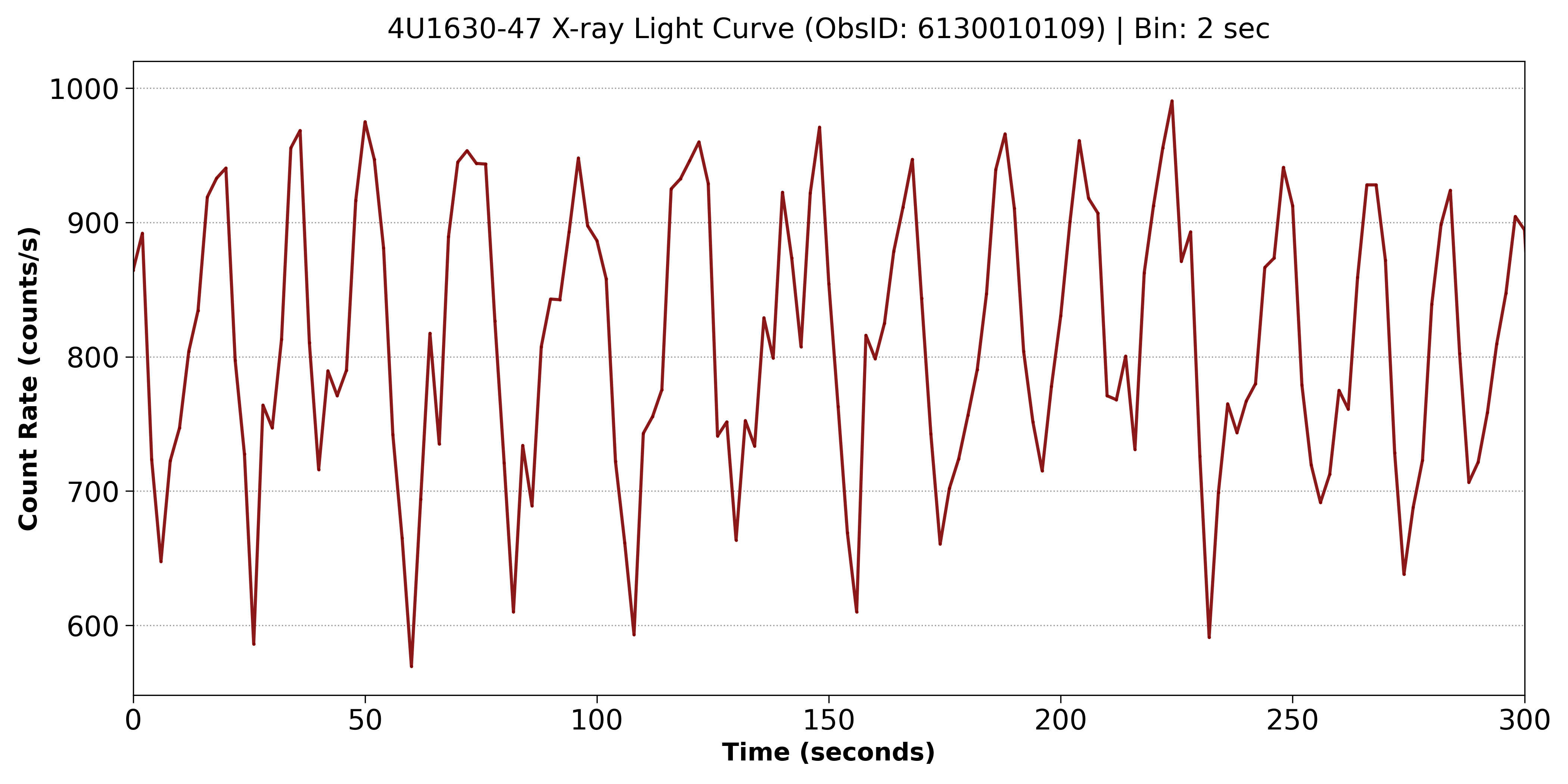}
    \caption{Lightcurve of the source 4U 1630-47, showing heartbeat states in NICER observation recorded on 2023 March 27 with obsID 6130010109. The lightcurve has been obtained using a binning interval of 2 seconds. The period of these oscillations are about 20-25 secs, cf. \cite{fan2025}.
    }
    \label{fig:4U_lightcurve}
  \end{figure}

\begin{figure}
  \centering
    \includegraphics[width=0.8\textwidth]{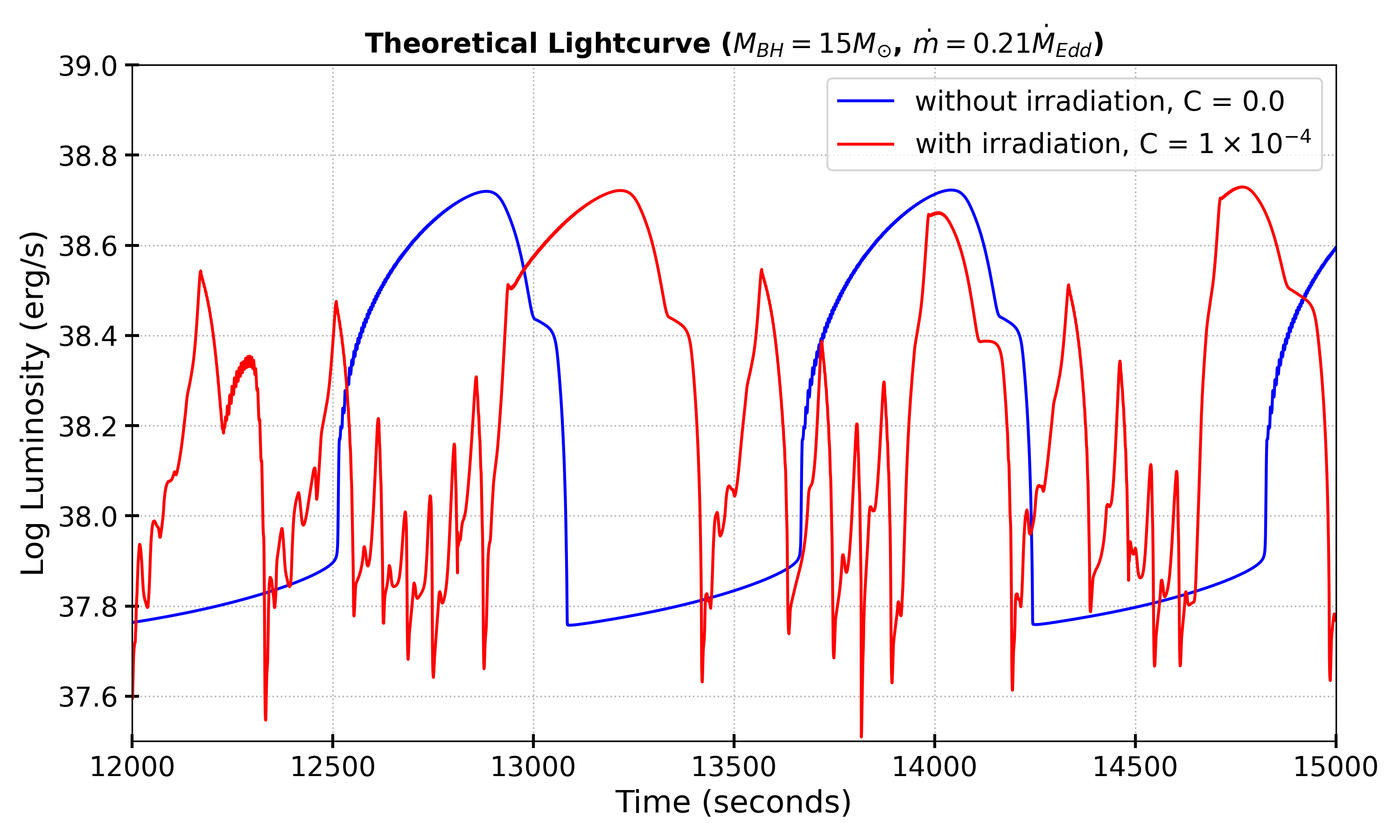}
    \caption{Theoretical lightcurves produced from GLADIS code including the effect of irradiation from the central compact region onto the disk around a black hole (red curve). For comparison, the non-irradiated disk evolution is shown (blue curve). The BH mass is 15 $M_{\odot}$, accretion rate is $\dot M = 0.21 \dot M_{\rm Edd}$, viscosity parameter is $\alpha=0.01$.}
    \label{fig:gladis_irra}
  \end{figure}

The temperature of the disk is changing due to additional heating from the central source, which is scaled with the local accretion rate and subject to geometrical effects.
 
 \begin{equation}
                T_c^4 = \frac{3}{8} \tau_{\rm tot} T_{\rm eff}^4 + T_{\rm irr}^4 \; ; \; T_{\rm irr}^4 = C \frac{\dot{M} c^2}{4 \pi R^2} 
                \end{equation}
In the above formula, 
the factor C includes the geometrical factor $\cos(\theta)$,
            \begin{equation}
                C = \eta (1 - \epsilon)\cos(\theta)
                \label{eq:C_dubus}
            \end{equation}
and $\eta$ denotes the X-ray efficiency, while $1 - \epsilon$ shows the fraction of X-ray absorbed in the optically thick layers of the disk. 
In Figure \ref{fig:geom_irr} we present a sketch of the geometry sketch adopted in the new version of our GLADIS code (under development) which accounts for the disk irradiation by the central source.
The geometrical factor $\cos(\theta)$ in Eq. \ref{eq:C_dubus} is defined as:
            \begin{equation}
                  \cos(\theta) = \frac{\frac{d H_{\rm irr}}{d R} - \frac{\Delta H_{\rm irr}}{\Delta R}}{\sqrt{(1 + (\frac{\Delta H_{\rm irr}}{\Delta R})^2)(1 + (\frac{d H_{\rm irr}}{d R})^2)}}
            \end{equation}
where $\Delta H_{\rm irr} = H-H_{\rm BL}$, and $\Delta R = R-R_{\rm BL}$ as marked in Fig. \ref{fig:geom_irr} and radial derivatives are computed during time dependent simulation, when the disk thickness is oscillating
(note that it is slightly different from the prescription of \cite{Dubus1999}. Our goal is to model observations such as those shown in Fig.
\ref{fig:4U_lightcurve} where we present the lightcurve of 4U1630-47 as observed by NICER on 2023 March 27 (cf. \cite{fan2025}). A preliminary result from the new version of GLADIS is shown in Figure \ref{fig:gladis_irra}, where it is evident that the extra role of irradiation (red line) alters the regular accretion cycle which GLADIS calculates in its absence (blue line). The irradiation appears to introduce a faster cycle on top of the main one, due to its effect on the local instability/mass accretion rate.

\subsection{Intermediate mass and supermassive BHs. Quasar duty cycles}

RPI instability can reproduce the duration, period, and amplitude of the outbursts in the intermediate-mass black hole, HLX-1  \cite{grzedzielski2017a}  as well as it can be applied to the AGN phenomenology, as shown in \cite{grzedzielski2017b}, who investigated the importance of atomic processes at lower temperatures.
With GLADIS code and a modified prescription of the viscous dissipation, the stabilizing effect of the iron opacity bump was examined and proven to be limited.

Quasars, in order to grow their central black hole masses should be active for $10^8$ years. However, there are arguments that this growth is not steady but it has an intermittent character, with timescales of the individual episodes of $\sim 10^5$ years. The observational evidence is based on the size of the photo-ionized region in a typical quasar \cite{Schawinski2015}. Another method to measure these timescales, giving similar results, is based on age estimates of the radio load sources and uses the jet size (kinetic method) or estimate of the synchrotron cooling timescale \cite{czerny2009}. Observations suggest also much more rapid changes in a given source, on timescales of years, stationary or propagating but they are most likely explained as re-collimation shocks \cite{lister2009}.  

Radiation pressure instabilities are expected to be present in AGN as the disks there are dominated by the radiation pressure. Models of such phenomenon \cite{czerny2009,grzedzielski2017a} imply the timescales of 100 to $10^5$ years, depending on the black hole mass and accretion rate. However, the details depend essentially on the description of the viscosity, as discussed by \cite{JaniukCzerny2011} (see \cite{grzedzielski2017a} for a more general treatment). Both the period, and particularly the outburst amplitude, are affected. There are also possible phenomena stabilizing the disk such as the presence of the warm corona, or winds. The statistical comparison of the models with the data is not simple as there are a number of mechanisms which should be included. Observational constraints on the ionized region can simply reflect the large density of the interstellar medium and too low jet or radiation power to expand further the ionized region, which then may not correspond to the source age.  

Theoretical approach to determine whether RPI mechanism indeed works are rather difficult, so comparing simple models to the observational data seems a much more promising path. There were attempts to explain the phenomenon of the rapid (timescales of a few years)  changes in the AGN appearance in sources known as Changing-Look AGN (CL AGN) \citep{sniegowska2020} but such short timescales in a roughly standard model can be only achieved in the case of a very small outer radius of the accretion disk \citep{sniegowska2023}. This in turn can be only possible if the disk is formed in the tidal disruption event. Alternatively, since most of the CL AGN accrete at a few per cent Eddington rate \citep{panda2024}, it may imply that the instability couples to the disruption of the inner flow, or is regulated by magnetic pressure, in a way which is not yet incorporated in the global modeling (see \cite{pan2022}).

\section{Open Questions}

{\bf State-of-the-art numerical simulations.}
Radiation-MHD simulations initiated by \cite{ohsuga2011} and recently presented e.g. by \cite{jiang2019} are the powerful tool that includes MRI transport, magnetic field evolution, vertical structure and radiation. Still, they resolve only local thermal timescales and due to 
limited duration cannot recover
full long-term cycles.
Technical challenges are met with the code's stiffness, and computations cost of radiation coupling. 
Magnetic pressure seems to be key factor that can modify or stabilize expected instability. 
The discrepancies between 1D long-term predictions and short-term multi-D outcomes reveal that
local boxes can be thermally unstable, but other results show magnetically supported stability 
(see S. Hirose review in this Proceeding).

Also, although the local MHD simulation postulates the stabilizing influence of the atomic processes, only the global time-dependent model can reveal the global disk stability range estimation. This is due to the global diffusive nature of those processes.

{\bf Outlook and future directions.}
New observations with time-domain facilities, and multi-wavelength
coordination are essential to obtain improved limit cycles and deterministic chaos metrics in wide class of astrophysical sources with accreting compact stars.
Theoretical advances should encompass: longer duration and multi D simulations, reduced-order informed
models, and coupling to other parts of accretion flows, such as boundary layers, coronas, winds and jets.
Cross-disciplinary tools that may help the advancements in the field are powerful Machine Learning tools for pattern recognition, and studies of other dynamical
systems for better interpretation.

\bigskip

{\bf Acknowledgments}
We thank Lukasz Klepczarek for technical support and discussions.
We acknowledge Poland’s high-performance Infrastructure PLGrid through 
HPC Centers: ACK Cyfronet AGH for providing computer facilities
and support within computational grant no. PLG/2025/018232 and Warsaw ICM within grant GCT under allocation g101-2281.
This work was supported by the
Polish National Science Center grant 2023/50/A/ST9/00527.

\end{document}